\documentclass[useAMS,usenatbib]{mn2e}
\input{psfig.sty}
\def\kms{\,{\rm km\,s^{-1}}}

\def\msun{\,{\rm M_\odot}}

\def\etal{{et al.\ }}

\newcommand{\intd}{{\rm int}}

\newcommand\beq{\begin{equation}}
\newcommand\eeq{\end{equation}}
\newcommand{\ba}{\begin{eqnarray}}
\newcommand{\ea}{\end{eqnarray}}
\def\spose#1{\hbox to 0pt{#1\hss}}
\def\lta{\mathrel{\spose{\lower 3pt\hbox{$\mathchar"218$}}
      \raise 2.0pt\hbox{$\mathchar"13C$}}}
\def\gta{\mathrel{\spose{\lower 3pt\hbox{$\mathchar"218$}}
      \raise 2.0pt\hbox{$\mathchar"13E$}}}

\title[Dynamical evolution and observable signatures  of IMBHs]
{Dynamical evolution of intermediate mass black holes and their observable
signatures in the nearby Universe}

\author[Marta Volonteri and Rosalba Perna]
{Marta Volonteri$^{1,2}$\thanks{E-mail: marta@ucolick.org (MV); rosalba@jilau1.colorado.edu (RP)} 
and Rosalba Perna$^{3}$\\
$^{1}$Department of Astronomy \& Astrophysics, University of 
California, Santa Cruz, CA 95064, USA\\
$^{2}$Institute of Astronomy, University of Cambridge, Madingley Road, Cambridge CB3 0HA, UK\\
$^{3}$Department of Astrophysical and Planetary Sciences,
University of Colorado at Boulder, 440 UCB, Boulder, CO 80309, USA}

\begin{document}

\date{}

\pagerange{\pageref{firstpage}--\pageref{lastpage}} \pubyear{2004}

\maketitle

\label{firstpage}

\begin{abstract}
We investigate the consequences of a model of the assembly and growth of massive black
holes (MBHs) from primordial seeds, remnants of the first generation of stars
in a hierarchical structure formation scenario. Our model traces the build-up of MBHs 
from an early epoch, and follows the merger history of dark matter
halos and their associated holes via Monte Carlo realizations of the
merger hierarchy from early times to the present time in a $\Lambda$CDM
cosmology.  The sequence of minor and major mergers
experienced by galactic halos in their hierarchical growth affects the 
merger history of MBHs embedded in their nuclei. So, if the formation route for the 
assembly of SMBHs dates back to the early universe, a large number of BH interactions 
is inevitable.  Binary black holes coalescence timescales can be long enough for a third 
BH to fall in and interact with the central binary. These BH triple interactions lead 
typically to the final expulsion of one of the three bodies and to the recoil of the binary. 
Also, asymmetric emission of gravitational waves in the last stages of the black hole merging 
can give a recoil velocity to the centre of mass of the coalescing binary.
This scenario leads to the prediction of a 
population of intermediate mass BHs (IMBHs) wandering in galaxy halos at the present epoch. 
We compute the luminosity distribution produced by
these IMBHs accreting from their circumstellar medium.
We find that in a Milky Way-sized galaxy they are unable to
account for sources with luminosities $\ga 10^{39}$ erg/s unless
they carry a baryonic remnant from which they are able
to accrete for a long time. We also find that, for typical
spiral galaxies, the bright end of the point source
distribution correlates with the mass of the galaxy, and
the most luminous sources are expected to be found
in the disk. 
\end{abstract}


\section{Introduction}
Observations of the centres of nearby galaxies indicate that most host a supermassive black 
hole (Magorrian \etal 1998; Ferrarese \& Merritt 2000; 
Gebhardt \etal 2000); supermassive black holes (SMBHs) seem 
to be unequivocally linked to galactic bulges, their masses scaling with the bulge luminosity 
and stellar velocity dispersion.  In cold dark matter (CDM) `bottom-up' cosmogonies, 
mergers of galaxies are a central part of the galaxy formation process in the hierarchical 
structure formation picture, with major mergers (i.e. those between comparable-mass systems)
expected to result in the formation of elliptical galaxies (Barnes 1988).

Volonteri, Haardt \& Madau 2003 (VHM03) and Volonteri, Madau, \& Haardt 2003 (VMH03)
have assessed a model 
for the assembly of SMBHS at the centre of galaxies that trace their hierarchical 
build-up,  assuming that the first
`seed' black holes (BHs) had intermediate masses, $m_\bullet\approx 150\,\msun$,
and formed in (mini)halos collapsing at $z\sim 20$ from high-$\sigma$ density 
fluctuations. These pregalactic holes evolve in a hierarchical fashion, following the merger 
history  of their host halos.  During a merger event BHs approach each other owing to 
dynamical friction, and form a binary system. Stellar dynamical processes drive the binary to 
harden and eventually coalesce.

The lifetime of BH binaries can be long enough (Begelman, Blandford \& Rees; 
Quinlan \& Hernquist 1997; Milosavljevic \& Merritt 2001; 
Yu 2002) that following another galactic merger a third BH can fall in and disturb the 
evolution of the central system (VHM03).
The three BHs are likely to undergo a complicated resonance scattering interaction, leading to 
the final expulsion of one of the three bodies (`gravitational slingshot') and to the recoil of 
the binary. Any slingshot, in addition, modifies the binding energy of the binary, typically
creating more tightly bound systems (VHM03).

Another interesting gravitational interaction between black holes happens
during the last stage of coalescence, when the leading physical 
process for the binary evolution becomes the emission of gravitational waves. 
If the system is not symmetric (e.g. BHs have unequal masses or spins) there would be a recoil due to 
the non-zero net linear momentum carried away by gravitational waves in the coalescence 
(`gravitational rocket'). The coalesced binary would then be displaced from the galactic 
centre, leaving straightaway the host halo or sinking back to the centre owing to dynamical 
friction. 

Historically, the possibility of these gravitational interactions between BHs has been envisaged even
before the CDM-hierarchical picture has become popular. Slingshots have been proposed,
as a contrast to the unified -twin beam- theory (e.g. Blandford \& Rees 1974)
to explain the nature of double radio sources (Saslaw, Valtonen \& Aarseth 1974; 
Mikkola \& Valtonen 1990; Valtonen \& Hein{\" a}m{\" a}ki 2000): ejected BHs can produce 
extended radio lobes by interacting with the intergalactic medium in the areas of the lobes. 

Slingshots and rockets were both involved 
when trying to explain the nature of galactic dark halos as composed of BHs 
(Hut \& Rees 1992; Xu \& Ostriker 1994). 
Both processes basically give BHs a recoil velocity, that can even exceed the escape 
velocity from the host halo and spread BHs outside galactic nuclei (VHM03). 
So, the outlined scheme predicts, along nuclear SMBHs 
hosted in galaxy bulges, a number of {\it wandering BHs} that are 
the result of rockets, slingshots and mergers with a dynamical friction timescale longer 
than the Hubble time, so at $z=0$ the BHs are still on their way to the galactic centre. 
Note that most of the wandering BHs masses are in an intermediate range between stellar mass 
BHs and supermassive ones.

In this paper, we improve the modelling of the dynamical interactions between galactic
halos and between BHs and their host halos. In particular, 
we include the lengthening of the dynamical 
friction  timescale of infalling satellites due to tidal interactions with the potential 
of the primary halo, and follow the orbital decay of every single satellite, determining 
its position within the primary at any given time. We also trace step-by-step the 
trajectories of BHs ejected from galaxy centres following an energetic BH interaction;
this enables us to keep 
track of the position and velocity of all wandering IMBHs as a function of cosmic time.

The population of wandering IMBHs in local galaxies that our model
predicts will have some observational consequences. In particular, due
to accretion from their surrounding interstellar medium (ISM), these
BHs will appear as bright X-ray sources. We compute here the
luminosity distribution expected from such sources, and relate it to
current observations. In particular, we discuss whether and under
what conditions these IMBHs can make up a substantial fraction of a
population of ultra-luminous sources (ULXs), characterised by
luminosities $L_X\ga 10^{39}$ erg/s, which has been recently
identified in large {\em Chandra} surveys (e.g. Colbert et al. 2004).  The properties of
this population as a whole (see Fabbiano 1988 for a review)
hint that we might be actually observing an
heterogeneous population, where super-Eddington luminosities could for
example be due to a photon-bubble instability (Begelman 2002) or
could be apparent due to strong beaming (King et al. 2001). A
contribution to the high-end tail of the X-ray point source luminosity
distribution could also be due to young millisecond pulsars (Perna \&
Stella 2004).  However, accreting IMBHs constitute an obvious
candidate for the ULXs, as it has been often suggested (e.g. Colbert
\& Mushotzky 1999; Miller \& Hamilton 2002; Cropper et
al. 2004). Part of our study here is therefore a detailed
investigation of the present-day observational properties of the
population of IMBHs formed in mini-halos at high redshifts and evolved
during the process of galaxy formation.

The paper is organised as follows: in \S2, we summarise the main features
of our cosmological scenario of BH growth; the dynamical evolution of
SMBH binaries is discussed in \S3. The dynamical
evolution of satellites BHs is studied in \S4, while \S5 describes the 
accretion model. \S6 is dedicated to the observational consequences of these IMBHs 
in present-day galaxies and \S7 discusses our results in the context of previous work 
in the field, and finally, in \S8, we summarise our work. 

\section[Assembly and mergers of SMBH\lowercase{s}]{Assembly and mergers of SMBH\lowercase{s}}
We briefly summarise here the main features of our scenario for the
hierarchical growth of SMBHs in a $\Lambda$CDM cosmology (see VHM03 for a 
thorough discussion). In our model, pregalactic `seed' holes form with 
intermediate masses ($m_\bullet=150\,\msun$) in (mini)halos 
collapsing at $z=20$ from rare 3.5-$\sigma$ peaks of the primordial density 
field (VHM03; Madau \& Rees 2001). The assumed `bias' assures that 
almost all halos above $10^{11}\,\msun$ actually host a BH at all epochs. 

The merger history of dark matter halos and associated BHs is followed by cosmological 
Monte Carlo realizations of the merger hierarchy from early times until the present in a 
$\Lambda$CDM cosmology. The dynamical evolution of the BH population is followed in detail with
a semi-analytical technique. In this paper, we focus on the properties of two sets of halos
with $M=2\times10^{12}\,\msun$ and $M=10^{13}\,\msun$ respectively, 
creating 30 merger trees
of the former and 10 of the latter and averaging the results. 
For the smaller halo, we will also consider a case in which seed holes are more numerous
and populate the 3-$\sigma$ peaks instead. 

In order to follow the history of halos and massive black holes, 
we adopt a two-component model for galaxy halos. The dark matter is distributed 
according to a NFW profile (Navarro \etal 1997). During the merger of two halo$+$BH 
systems of comparable masses, dynamical friction against the dark matter background  
drags in the satellite hole towards the centre of the newly merged system,
leading to the formation of a bound BH binary. When two halos of mass $M$ and $M_s$ merge, 
the `satellite' (less massive) progenitor (mass $M_s$) is assumed to sink to the 
centre of the more massive pre-existing system on the dynamical 
friction (against the dark matter background) timescale, which depends on the orbital 
parameters of the infalling satellite, which we take from van den Bosch \etal (1999).

The fate of the infalling halo has been investigated recently (Taffoni et al. 2003) 
in a NFW profile. Dynamical friction appears to be very efficient for 'major' mergers 
with (total) mass ratio of the progenitors, $P=M_s/M>0.1$. In each major merger the more 
massive hole accretes at the Eddington rate a gas mass that scales with the fifth power 
of the circular velocity of the host halo: the normalisation is fixed a posteriori in order
to reproduce the observed local $m_{\rm BH}-\sigma_*$ relation (Ferrarese
2002) and the observed 
luminosity function of optically-selected quasars in the redshift range $1<z<5$.

Satellites of intermediate mass ($0.01 M_{\rm h}<M_{\rm
s,0}<0.1 M_{\rm h}$) suffer severe mass losses by the tidal
perturbations induced by the gravitational field of the primary halo.
Tidal perturbations cause the progressive stripping of the satellite and, 
as the magnitude of the frictional drag depends on the mass of the satellite, 
this progressive mass loss increases the decay time. 
After a Hubble time satellites have typical masses $\sim 1-10\%$ of their 
initial mass.
The lightest satellites ($P<0.01$) seem to be almost 
unaffected by orbital decay, so they survive and keep orbiting on rather circular, 
peripheral orbits. 

\begin{figure}
\psfig{file=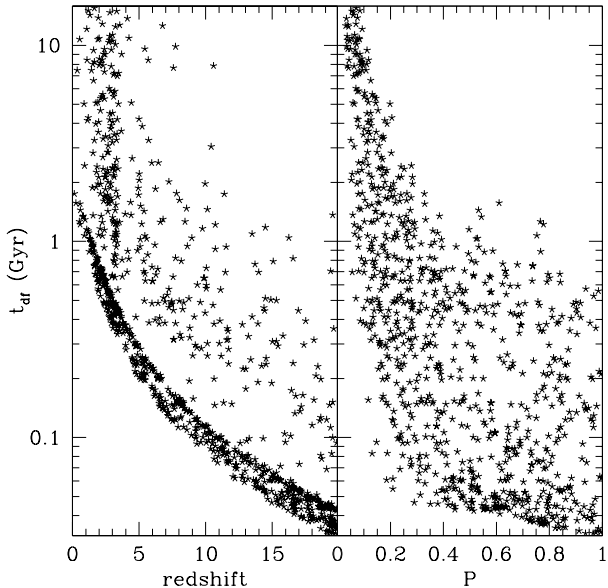,width=84mm}
\caption{Dynamical friction timescale of the halos hosting massive black holes as a function of redshift 
({\it left panel}) and mass ratio  ({\it right panel}). 
Note at low redshift ($z<5$) the sharp division in timescale for major and minor mergers, 
the former characterised by timescales shorter than 2 Gyrs, the latter can be longer 
than the Hubble time.}
\label{figtdf}
\end{figure}

We adopt the relations suggested by Taffoni \etal for the orbital 
decay of satellites in a NFW halo.
Figure \ref{figtdf} shows the evolution of the dynamical friction timescales with 
redshift and the dependence on the mass ratio for halos both hosting massive black holes.

\section{Black hole binary dynamical evolution}

The subsequent evolution of a BH binary is determined by the initial 
central stellar distribution. The binary will initially shrink by dynamical friction from 
distant stars acting on each BH individually. But as the binary separation 
decays, the effectiveness of dynamical friction slowly declines because distant encounters 
perturb only the binary centre of mass but not its semi-major axis. The BH pair 
then hardens via three-body interactions, i.e., by capturing and ejecting at 
much higher velocities the stars passing by within a distance of order 
the semi-major axis of the binary. The merger timescale is computed adopting a simple 
semi-analytical scheme that qualitatively reproduces the evolution observed
in N-body simulations (Merritt 2000; VHM03, VMH03). 
The binary evolution is slowed down due to the declining stellar density, with a 
hardening time that becomes increasingly long as the binary shrinks. 
If the hardening continues sufficiently far, gravitational radiation losses finally take over, 
and the two BHs coalesce in less than a Hubble time.

We have adopted a simplified description of the BH coalescence process, ignoring for instance 
the depopulation of the loss cone (e.g. Yu 2002) and the loss of orbital angular momentum to 
a gaseous disc (Armitage \& Natarajan 2002, Escala \etal 2004).

\subsection{Gravitational Rocket}
When the binary semi-major axis is such that the timescale for hardening by stellar
scatterings equals the timescale for shrinking by emission of gravitational waves, the 
latter process takes over.  At this point, if the holes
have unequal masses, or spin, the binary recoils due to the non-zero net linear momentum
carried away by gravitational waves in the coalescence (`gravitational rocket'). 

The recoil velocity has still large uncertainties, and we adopt here  
the results from recent calculations by Favata et al. (2004). They estimate the recoil velocity as a 
function of the binary mass ratio $q=m_2/m_1$ (with BH masses $m_1\ge m_2$) and the spin parameter 
during final plunge and coalescence. We adopt here the expression they suggest for estimating 
'upper limits' for the recoil, converting it into estimates of the binaries 
recoil velocity, $v_{\rm CM}$, following the procedure described in Merritt et al. (2004).

If the recoil velocity of the coalescing binary exceeds the escape 
speed $v_{\rm esc}$, the holes will leave the galaxy altogether. 
If instead $v_{\rm CM}<v_{\rm esc}$, we assume that the binary is kicked out to a radius 
$r_{\rm max}$ such that the total energy is conserved, i.e. we solve for $r_{\rm max}$ 
the equation
$(m_1+m_2)\,v_{\rm CM}^2=2(|\phi_0|-|\phi(r_{\rm max})|)$, 
where $\phi_0$ is the galactic potential at the initial location of the binary 
and $\phi(r_{\rm max})$ is the potential at the apocentre.

\subsection{Gravitational slingshot}
The dynamical evolution of SMBH binaries may be disturbed by a third  
incoming BH, if another major merger takes place before the pre-existing binary has 
had time to coalesce or a wandering BH can reach the centre owing to dynamical friction
(e.g. Hut \& Rees 1992; Xu \& Ostriker 1994). 
If the incoming hole reaches the sphere of influence of the central binary, the three BHs are 
likely to undergo a complicated resonance scattering interaction, leading to the final expulsion 
of one, typically the lightest, of the three bodies (gravitational slingshot). 

Typically an encounter with an intruder of mass $m_\intd$ smaller than both 
binary members leads to a scattering event, where the binary recoils by momentum 
conservation and the incoming lighter BH is ejected from the galaxy nucleus. 
By contrast, when the intruder is more massive than one or both  
binary components, the probability of an exchange is extremely high: the incoming hole
becomes the member of a new binary, and the lightest BH of the original pair gets
ejected (Hills \& Fullerton 1980). 

All triple interactions are followed along the merger tree, 
adopting the same scheme described in VHM03.

In all cases we have estimated the recoil velocities conserving the binary energy and momentum
and estimated the apocentre $r_{\rm max}$ solving the equation
$(m_1+m_2)\,v_{\rm CM}^2=2(|\phi_0|-|\phi(r_{\rm max})|)$.
Dynamical friction timescales have been calculated by the 
Chandrasekhar dynamical friction timescale for circular orbits, so they are presumably
upper limits to these timescales (thus giving an upper limit on the number of wandering BHs), 
as the ejection probably occurs on eccentric orbits.  

\begin{figure}
\psfig{file=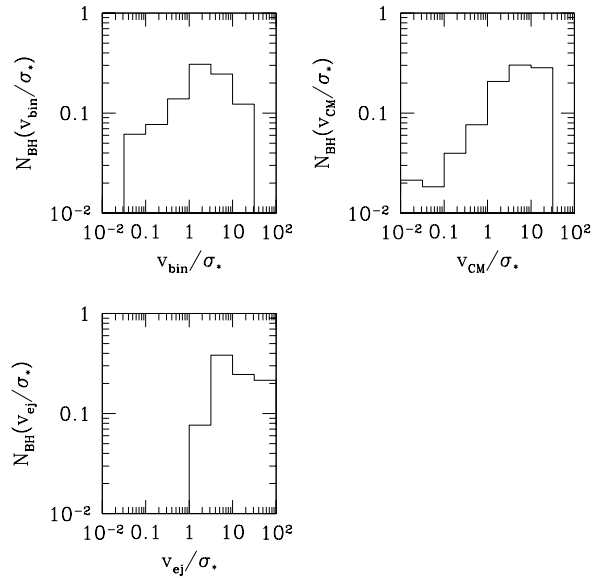,width=84mm}
\caption{Distribution of BH recoil velocities (in units of the stellar velocity dispersion). 
Binaries, either following a triple interaction (upper left panel) or a rocket (upper right panel)
, have a broad velocity 
distribution, while almost all single BHs have high recoil velocities, typically larger than 
the escape velocity.}
\label{vel}
\end{figure}

In figure \ref{vel} the distribution of BH recoil velocities is plotted in units of the stellar 
velocity dispersion, ranging from $5 \kms$ to $200 \kms$ depending on redshift and halo mass. 
As a rule of thumb, if the recoil velocity is smaller than the
velocity dispersion, the BH remains in the proximity of the centre and the orbital decay
is fast (Madau \& Quataert 2004).  Figure \ref{vel} shows that the ejection of all three 
BHs is therefore not a likely outcome under the assumptions of our BH growth model 
(see VHM03 for a thorough discussion).

Clearly, most of the single holes either escape their 
host halos or are slung to the periphery of the galaxy with consequently long dynamical 
friction timescales. Most of the binaries recoil instead within the core and fall back to 
the centre soon afterwards, with $t_{\rm df}<0.01\,$Gyr. 

Note that triple interactions are rare events, in the 
whole history of a MW-sized halo with its satellites, on average only $\sim 3$ slingshots
happen, in the case of seed holes in 3.5-$\sigma$ peaks.

\section{Dynamical evolution of off-nuclear black holes}
The assumptions underlying our scenario lead to the prediction of a population of massive BHs 
{\it wandering} in galaxy halos, due to the processes described above.

During the evolution and mass growth of DM halos, we track the position of all wandering BHs,
following the modifications of the galactic potential due to mergers.
In case of minor mergers, the potential varies slightly, so the wandering BHs orbits
evolve smoothly, while in case of major mergers the potential of both interacting systems
change abruptly, causing larger modifications to the BH orbits. Although numerical
simulations with resolution high enough to follow the dynamical evolution of intermediate 
mass BHs in galactic mergers are not available, we can rely on simulations studying the 
mixing of particles in major mergers (White 1980, Nipoti et al 2003), assuming that during 
a merger a BH follows the evolution of other particles. On average particles which were
at a given radius (normalised at the half mass radius) in one of the galaxies before the 
merger, are found at a similar radius (normalised to the new half mass radius) in the merger 
remnant. 

To track the orbital evolution of wandering BHs during galactic mergers we adopt the
following scheme:

\begin{itemize}

\item{MINOR MERGERS:} we let the wandering BHs in the more massive progenitor unperturbed, 
while the wandering BHs in the satellite galaxy are assumed to decay with their host. 

\item{MAJOR MERGERS:} each wandering BH outside the core is placed in the new halo at a 
distance conserving the same ratio with the virial radius as it had in its parent system. 
If a wandering BH is instead inside the core, it is assumed to reach the centre 
of the newly formed halo within the same merging timescale of the halos (see section 2).

\end{itemize}

\section{Detecting Intermediate-Mass Black Holes in local galaxies}
\subsection{Naked wandering BHs}
As our first model of IMBHs in present-day galaxies, we
assume that all BHs are \emph{naked}, i.e. that the BHs left over by minor 
mergers have been completely stripped of their envelope. We
embed all wandering BHs in a galactic disc/bulge/halo model (Springel \& White 1999), 
and determine the corresponding density of the medium in the surrounding of the BHs
at their respective positions.
We assume that a fraction $m_{\rm d}$ of the initial halo mass is
into a thin stellar disc with an exponential surface density, 
\beq
\rho_{\rm d}(R,z) = \frac{\Sigma_0}{2z_0} \exp\left(-\frac{R}{R_{\rm d}}\right) {\rm
sech}^2\left(\frac{z}{z_0}\right) .
\eeq
Here $M_{\rm d}=m_{\rm d} M$ is the total mass of the disc and
$R_{\rm d}$ is its scale radius. We adopt $m_{\rm d}=0.05$, while the
disc scale length is determined through the fitting formula provided in Shen et al. (2003) for 
the Petrosian half-light (or mass) radius:
\beq
R_{\rm d} = 0.17 M_{\rm d}^{0.14}  (1+M_{\rm d}/3.98\times10^{10})^{0.25}.
\eeq
Most spiral galaxies seem to be consistent with a constant vertical scale
length with a value of $z_0\simeq 0.2 R_{\rm d}$.
Bulges are modelled with a spherical Hernquist (1990) profile of the form
\beq
\rho_{\rm b}(r)=
\frac{M_{\rm b}}{2\pi}\frac{r_{\rm b}}{r(r_{\rm b}+r)^3} .
\eeq
In analogy to the treatment of the disc, we assume that the bulge mass is a fraction
$m_{\rm b}=0.02$ of the halo mass and that the bulge scale radius $r_{\rm b}$ is a
fraction $f_{\rm b}=0.1$ of that of the disc, i.e.\ $r_{\rm b}=f_{\rm
b} R_{\rm d} $. We have then assumed that the gas traces the baryons, the gas fraction 
(of the total baryonic mass) is $20\%$ in the disk and $4\%$ in bulge and halo 
(Bell \& de Jong 2000, Fukugita, Hogan \& Peebles 1998).

To model an elliptical galaxy (in the case $M=10^{13}\,\msun$) we have considered 
the Hernquist profile only, 
assuming that the baryonic content of the galaxy is a fraction $0.1$ of the halo mass.
For the elliptical, we have then assumed that the gas fraction is $1\%$(Mathews \& Brighenti 2003).

\subsection{Accretion model}

In this section, we discuss how we estimate the luminosities of
off-nuclear BHs resulting from the 
process of accretion from the surrounding medium. 

Let $\rho$ be the density of the medium in the surrounding of a BH,
and let $c_s$ be its sound speed. The accretion rate onto a BH 
of mass $M_{\rm BH}$ can be estimated using the Bondi-Hoyle formula
(Bondi \& Hoyle 1944),
\beq
\dot{M}_{\rm Bondi}=\frac{\lambda \,4\pi\, G^2\, 
M_{\rm BH}^2\,\rho}{(v^2+c_s^2)^{3/2}}\; ,
\label{eq:dotm}
\eeq  
where $v$ is the velocity of the BH with respect to the medium, and
$\lambda$ is a parameter on the order of 1. The BH radial velocities are
computed by integrating the Chandrasekar formula (Binney \& Tremaine 1987)      
as a function of the initial conditions, while the sound speed velocity,
$c_s\sim 10(T/10^4{\rm K})^{1/2}$ km/s, depends on the local temperature
at the BH location, which is discussed later in this section.

The bolometric luminosity of the BH can be written as
\beq
L_{\rm bol}=\eta \dot{M} c^2\;,
\label{eq:lbol}
\eeq
where $\eta$ represents the fraction of the accreted mass that is
radiated away. The nature of the accretion process, and the 
consequent value of $\eta$ in BHs is rather uncertain. Active
Galactic nuclei (AGNs) accrete through accretion discs with
an high efficiency ($\eta\sim 0.1$). Their luminosities are close
to their Eddington values. At the other end of the luminosity function,
are the SMBHs at the centres of our own and
nearby galaxies, whose luminosities can be as low as  $\sim 10^{-9}-10^{-8}$ 
of their Eddington values (e.g. Loewenstein et al. 2001). 
Low luminosities could be due either
to low efficiencies with $\dot{M}\sim \dot{M}_{\rm Bondi}$, or to
relatively high efficiencies with  $\dot{M}\ll \dot{M}_{\rm Bondi}$
(or also to a combination of both low accretion rates and low efficiencies). 
Both scenarios have been suggested in the literature.
Narayan \& Yi (1995a) found that, when the accretion rate decreases
below some critical value $\dot{m}\sim 0.1\alpha^2$ ($\alpha$ being
the viscosity parameter in the disc), the thin disc switches to
an Advection Dominated Accretion Flow, characterised by a low
radiative efficiency, due to the inability of the electrons
in the plasma to cool (see also Rees et al. 1982; Begelman \& Celotti 2004). 
On the other hand, an accretion rate at values below the Bondi rate
could be due to the presence of winds (or outflows; Narayan \& Yi 1995b;
Blandford \& Begelman 1999). This scenario appears to be supported
by observations of both accreting black holes (Loewenstein et al. 2001), as well
as old, accreting neutron stars (Perna et al. 2003).  

In this work, when considering a model for the X-ray luminosity from
accretion onto the IMBHs similar to that in SMBHs, we will adopt an empirical
approach, and simply assume that the X-ray luminosity is a fraction
$\eta_{\rm X}$ of the accretion luminosity, with $\eta_{\rm X}$ calibrated on
X-ray observations of SMBHs. This approach is similar to
the one adopted by Agol \& Kamionkowski (2002), in their calculations
of the X-ray luminosity from isolated, stellar-mass BHs in the Milky Way
(they parameterised their results in terms of the X-ray accretion
efficiency).

As a second model, we consider an optically thick, geometrically thin
accretion disc, as in the standard Shakura-Sunyaev (1973, SS) formulation.
Radiation is emitted locally with a blackbody spectrum of temperature
\beq
T(r) = \left(\frac{3GM_{\rm BH}\dot{M}}{8\pi r^3\sigma}\right)^{1/4}\left[
1-\left(\frac{r_{\rm in}}{r}\right)^{1/2}\right]^{1/4}\;. 
\label{temp}
\eeq
The emitted spectrum (per unit frequency) from the whole surface of the disc is given by
\beq
L_\nu= 2\pi \frac{h\nu}{c^2}\int_{r_{\rm in}}^{r_{\rm max}}\frac{r\, dr}
{e^{h\nu/k T(r)}-1}\;.
\label{eq:bb}
\eeq 
We take $R_{\rm in}=3 R_S$ and $R_{\rm max}=1000 R_S$, where $r_s=2GM_{\rm BH}/c^2$ is the Schwarzschild radius
of the BH; this yields an efficiency $\eta\sim0.06$. As described
in the previous sections,
our simulations allow us to identify whether the BHs are in the halo,
bulge or disc of the galaxy. For those in the halo and bulge, we
assume that the temperature is the virial temperature of the accreting
interstellar medium (ISM).  
This is $T_{\rm vir}=5.8\times 10^5$ K
for the galaxy of mass $2\times 10^{12}\msun$, and $T_{\rm vir}=2.3\times 10^{6}$ K 
for the galaxy of mass $10^{13}\msun$. 
In our Galaxy, we know that the disc is composed of several phases
(e.g. Bland-Hawthorn \& Reynolds 2000), 
from cold ($T\sim 100$ K), to warm ($T\sim 8000$ K), to hot
($T\sim 10^6$ K).  Modelling the various phases is beyond the scope of
this paper essentially because, while some information on these phases
is available for our Galaxy, no such studies have been possible for
other galaxies. Moreover, due to the process of radiative feedback
(Blaes, Warren \& Madau 1995), the gas in the vicinity of the BH will
be heated by the ionising radiation produced by accretion. 
The amount of heating and  cooling  depends on a number
 of factors, such as the density of the medium, its metallicity,                
the velocity of the BH, the accretion efficiency. For a medium                 
made of H and He only, the ratio between the total heating rate                
to the initial energy density of the gas at the accretion radius               
is $\sim 2\times 10^5\eta v_{40}^{-2}T_4^{-1}$ (Agol \& Kamionkowski           
2002), where $T_4=T_410^4$ K, and $v=v_{40}40$ km/s is the BH velocity.        
For BHs with velocities smaller than a few tens of km/s, heating is            
important even for very low accretion efficiencies. Once H and He are          
ionized, the opacity drops substantially, and the heating rate becomes         
much longer than the accretion rate, leaving the gas at a temperature          
of around $2\times 10^4$ K (Agol \& Kamionkowski 2002; Blaes, Warren \& Madau 1995). 
We will take this to be the temperature of the                    
accreting gas in the disc, assuming that, if there is a hot phase, this does not 
generally dominate.      Note that, in those cases where the velocity of the BH is sufficiently 
high that dynamical heating is not important (and therefore the gas is    
actually cooler than assumed), the velocity of the BH dominates over that      
of the sound speed in the expression for the accretion rate, hence making      
the precise value of the temperature of the gas unimportant for our calculations.                                                                   

For any BH, the density of the accreting material is determined
at its location in the galaxy according to the model described in \S 6.1.
For satellites, we also consider the possibility that they might accrete
from a remnant core as described in \S 5.4.

\section{Results and implications for ULXs}
At a given time, a galaxy not only contains off-centre BHs slingshot 
or rocketed, but can also have BHs, possibly embedded in the remnant of their original 
host, still on their way to the centre following a minor merger. 

\begin{figure}
\psfig{file=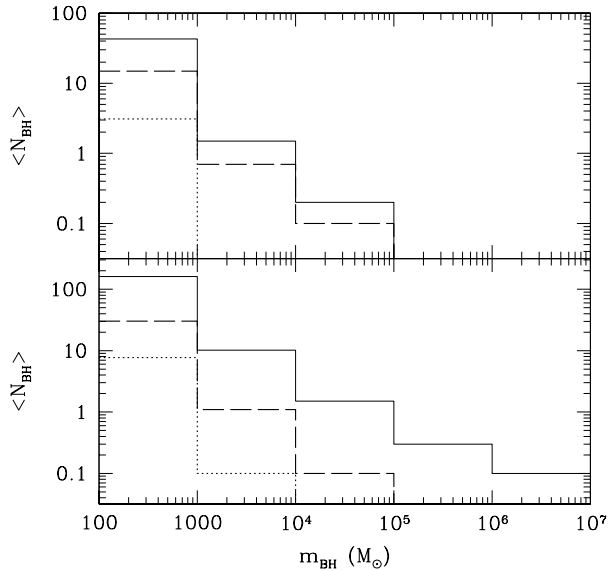,width=84mm}
\caption{ Mass function of wandering BHs at different redshifts, averaged over 
30 Monte Carlo realizations of a $M=2\times10^{12}\,\msun$ halo assuming that seed BHs form in 
3-$\sigma$ density peaks at z=20. {\it Upper panel}: BHs in 
satellite remnants, {\it lower panel}: BHs displaced from the centre due to slingshots or 
rockets. {\it Solid line}: z=0, {\it dashed line}: z=2, {\it dotted line}: z=5.}
\label{massfz}
\end{figure}

\begin{figure}
\psfig{file=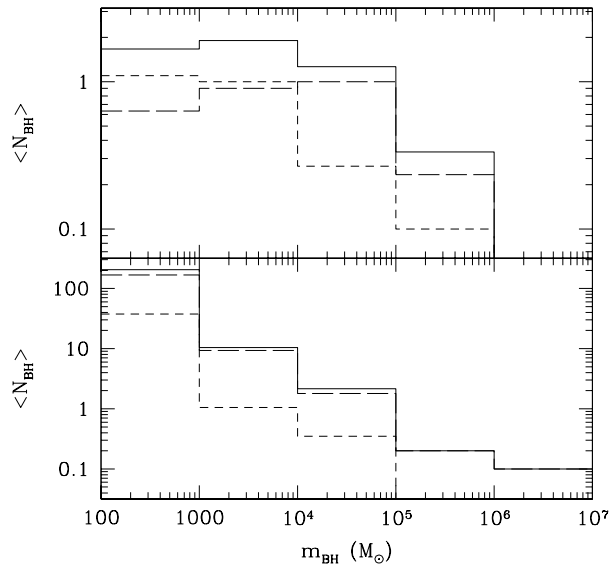,width=84mm}
\caption{Mass function of wandering BHs at z=0, averaged over 
30 Monte Carlo realizations of a  $M=2\times10^{12}\,\msun$halo. {\it Upper panel}: seed BHs
form in 3.5-$\sigma$ peaks, {\it lower panel}: seed BHs form in 3-$\sigma$ peaks. 
{\it Solid line}: total mass function, {\it long dashed line}: BHs displaced from the centre 
due to slingshots or rockets, {\it short dashed line}: BHs in satellite remnants.}
\label{massfz0}
\end{figure}

\begin{figure}
\psfig{file=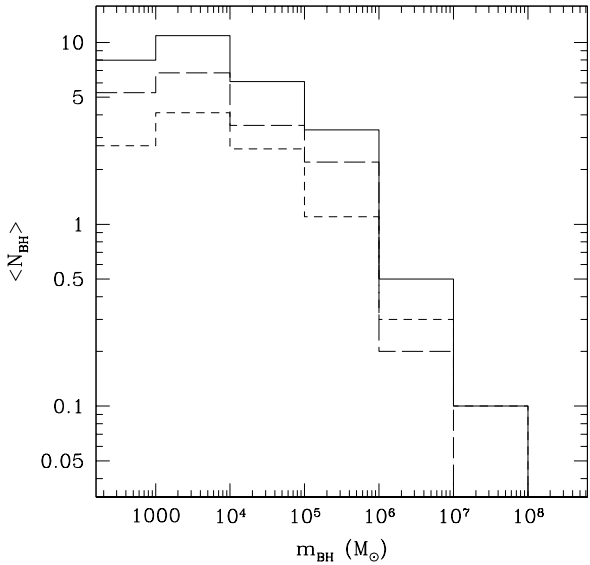,width=84mm}
\caption{ Mass function of wandering BHs at z=0, averaged over 
10 Monte Carlo realizations of a  $M=10^{13}\,\msun$halo.
{\it Solid line}: total mass function, {\it long dashed line}: BHs displaced from the centre 
due to slingshots or rockets, {\it short dashed line}: BHs in satellite remnants.}
\label{massfz0G}
\end{figure}

Figure \ref{massfz} shows the evolution with redshift of the mass function of wandering BHs in
a $M=2\times10^{12}\,\msun$ halo, assuming that seed holes form in 3-$\sigma$ density peaks 
at z=20. Figure \ref{massfz0} shows instead the mass function of wandering BHs in a
$M=2\times10^{12}\,\msun$ halo at $z=0$, for the case of seed holes in
3.5-$\sigma$ peaks and in 3-$\sigma$ peaks. Note that most of the
wandering BH masses are in an intermediate range between stellar mass
BHs and supermassive ones.  The upper end of the mass distribution of
IMBHs, as well as their total number, is an increasing function of
the galaxy mass. This can be seen by comparing Figure 4 (top panel)
and Figure \ref{massfz0G}, which shows the mass function of wandering
BHs in a $M=10^{13}\,\msun$ halo at $z=0$, for the case of seed
holes in 3.5-$\sigma$ peaks.

\begin{figure}
\psfig{file=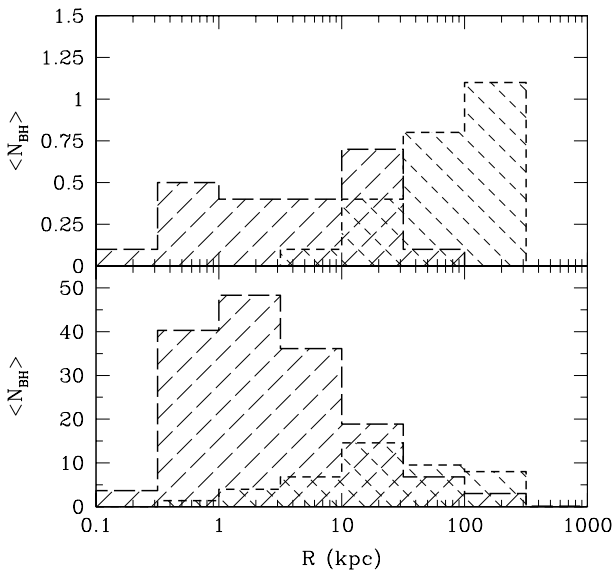,width=84mm}
\caption{ Radial distribution of wandering BHs at z=0, averaged over 
30 Monte Carlo realizations of a $M=2\times10^{12}\,\msun$ halo. {\it Upper panel}: seed BHs
form in 3.5-$\sigma$ peaks, {\it lower panel}: seed BHs form in 3-$\sigma$ peaks. 
{\it Long dashed line}: BHs displaced from the centre 
due to slingshots or rockets, {\it short dashed line}: BHs in satellite remnants.}
\label{rdistr}
\end{figure}

\begin{figure}
\psfig{file=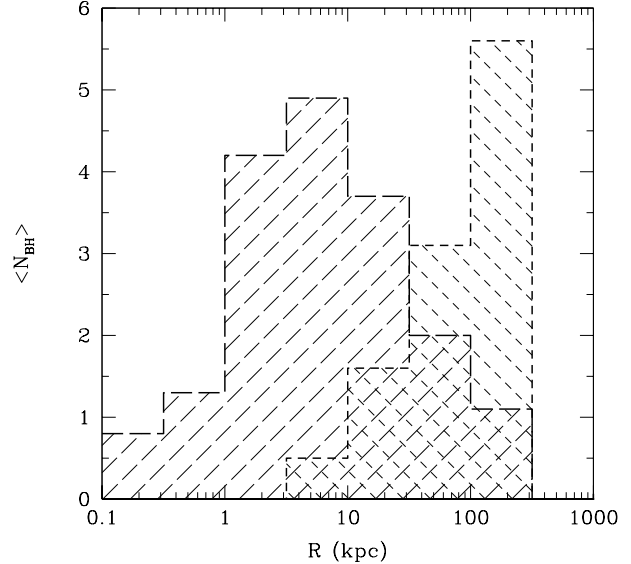,width=84mm}
\caption{ Radial distribution of wandering BHs at z=0, averaged over 
10 Monte Carlo realizations of a $M=10^{13}\,\msun$ halo. 
{\it Long dashed line}: BHs displaced from the centre 
due to slingshots or rockets, {\it short dashed line}: BHs in satellite remnants.}
\label{rdistrG}
\end{figure}

\begin{figure}
\psfig{file=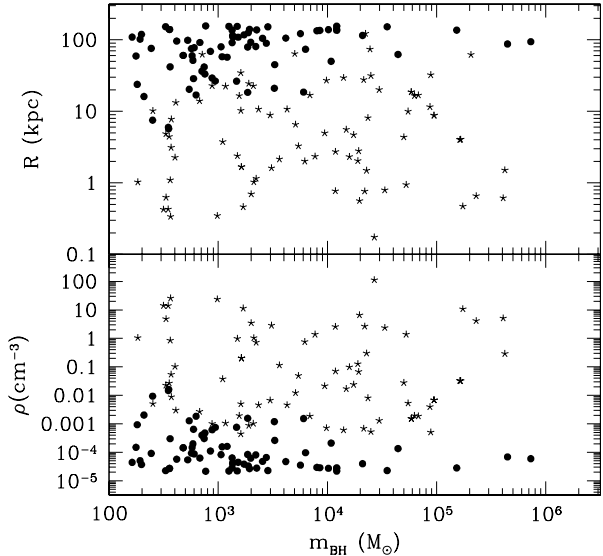,width=84mm}
\caption{Gas density ({\it lower panel}) and radial distance ({\it upper panel}) of BHs as 
a function of their mass, for the all 30 Monte Carlo realizations of a 
$M=2\times10^{12}\,\msun$ halo, with seed BHs in 3.5-$\sigma$ peaks. 
{\it Filled dots:} BHs in satellite remnants, {\it stars:} wandering BHs due to 
slingshots or rockets.}
\label{mr}
\end{figure}

Figures \ref{rdistr} and \ref{rdistrG} show the corresponding
radial distributions of these BHs. 
Typically, BHs in satellite remnants dwell in the outskirts of galaxies, while those 
caused by slingshots or rockets have a broader distribution and can be found even in the 
inner regions.
Interestingly, there is not a correlation between BH masses and radial distance 
(Figure \ref{mr}).  Naively, more massive BHs would be expected in inner regions, due to 
the inverse dependence of orbital decay timescale on the mass. On the other hand, 
the shortness of the orbital  decay timescale for the more massive BHs implies that very 
few of these are expected in the inner regions, because once they sink to the inner         
regions, they quickly sink further to the very center and merge with the       
central black hole or are re-ejected.   
Larger holes, however, appear only at later times, so that the time elapsed from their 
initial interaction with the galaxy is short. Figure  \ref{mr} shows also the density of 
gas surrounding the same BHs.

\begin{figure}
\psfig{file=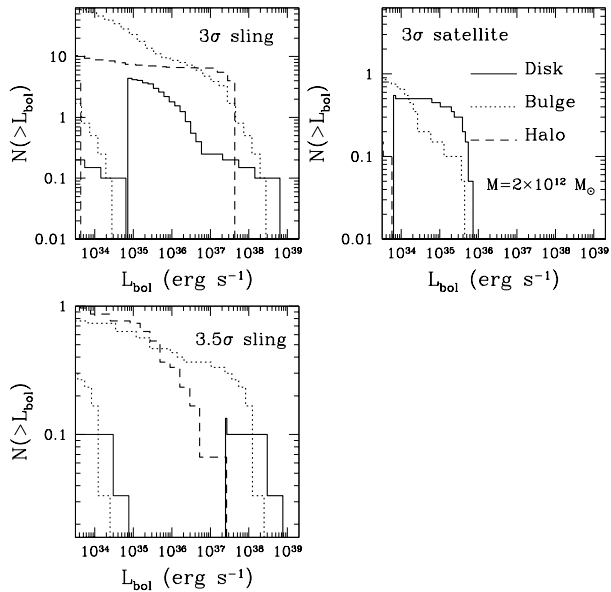,width=84mm}
\caption{ Cumulative distribution of the
bolometric luminosity for IMBHs accreting from the 
local ISM in a spiral galaxy of mass $M=2\times 10^{12}M_\odot$. Thick lines
refer to an accretion efficiency $\eta=0.1$, while thin lines are computed
with $\eta=10^{-5}$. The contribution from the three galaxy components
(disc, bulge and halo) is separately shown. The standard deviation among
the various Monte Carlo realizations at the upper end tail of the distribution
is 0.3 for the 3$\sigma$ sling case,
0.22 for the 3$\sigma$ satellites, and 0.18 for the 3.5$\sigma$ sling.}
\label{lbol1}
\end{figure}

\begin{figure}
\psfig{file=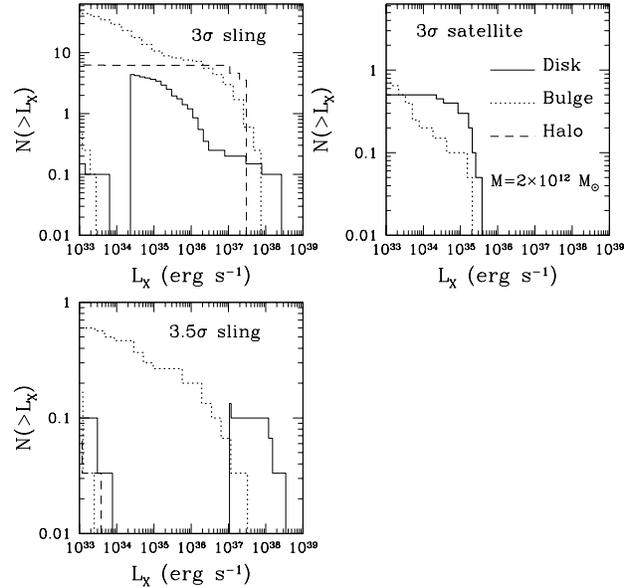,width=84mm}
\caption{ Cumulative distribution function for the X-ray luminosity 
(0.1-10 keV band) of IMBHs accreting from the local ISM through a SS-disc
(thick lines) and for an X-ray efficiency of $10^{-6}$ (thin lines). The galaxy, of mass
$2\times 10^{12}M_\odot$, is assumed to be a spiral, and the contribution
from the disc, bulge and halo is separately shown.}
\label{lx1}
\end{figure}

\begin{figure}
\psfig{file=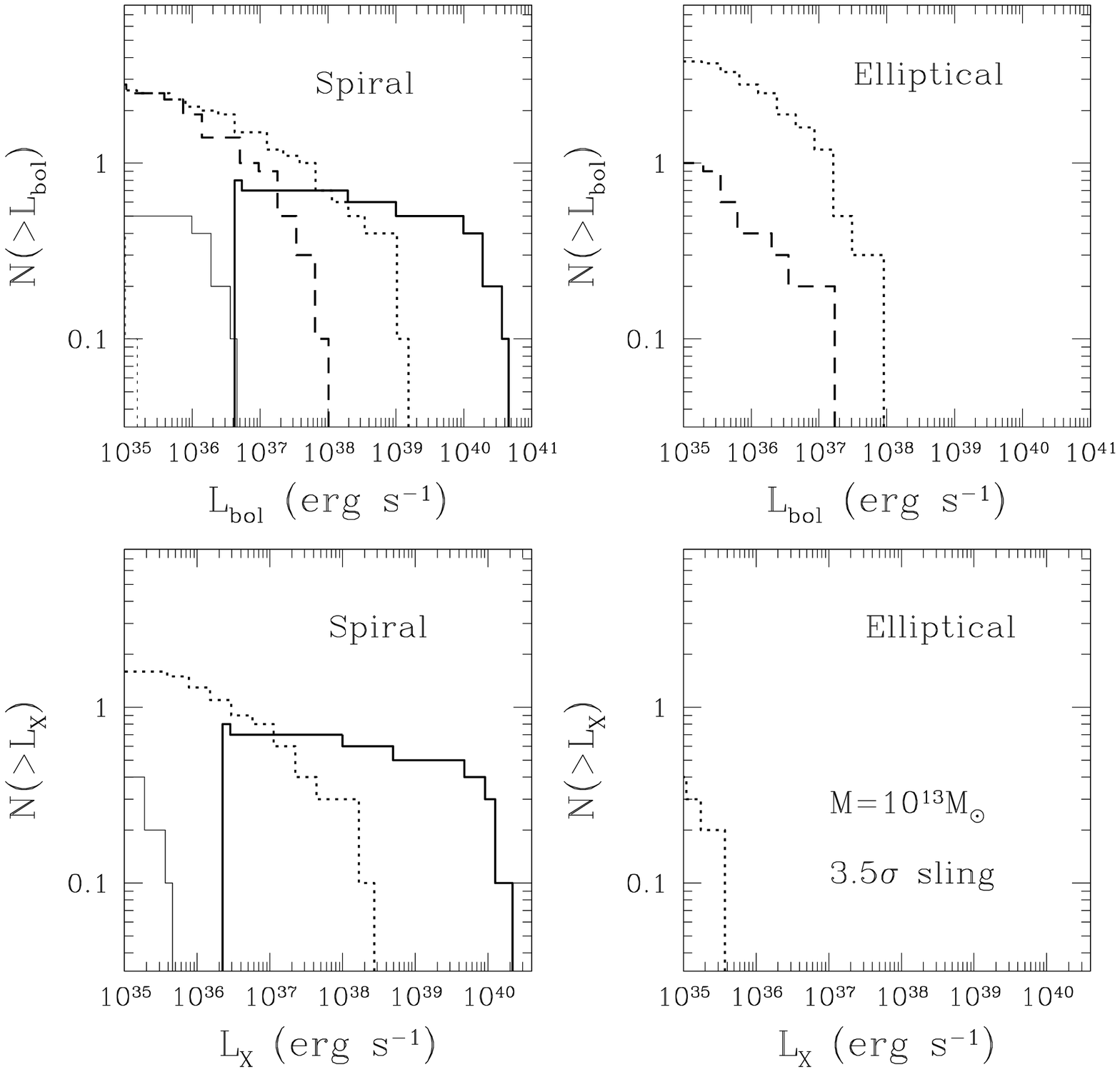,width=84mm}
\caption{ The upper panels show the bolometric luminosity distribution 
for a galaxy of mass $M=10^{13}M_\odot$ in a spiral (left) and elliptical
(right) model. Thick lines are computed assuming an efficiency of 0.1, while
thin lines of $10^{-5}$. For the same galaxy models, the bottom panels show 
the X-ray luminosity (0.1-10 keV band)
assuming that accretion proceeds through a SS-disc (thick lines)
and for an X-ray efficiency of $10^{-6}$ (thin lines). Line types
refer to the same galaxy components as in Figs. 9 and 10.
The standard deviation among
the various Monte Carlo realizations at the upper end tail of the distribution
is 0.31 for the Spiral and 0.48 for the Elliptical.}
\label{l2}
\end{figure}

Resulting luminosities are shown in Figures 9, 10 and 11. For the bolometric
luminosities, we show results for an efficiency $\eta=0.1$ (more
typical of AGNs) and for $\eta=10^{-5}$ (more
representative of the dormant SMBHs in local galaxies), while X-ray
luminosities are shown for the disc model and for an X-ray 
efficiency\footnote{The SMBH in the Galactic Centre has an X-ray efficiency of
$\sim 4\times 10^{-8}-2\times 10^{-6}$ in the 2-10 keV band
(Baganoff et al. 2001). For BHs at the
other end of the mass spectrum (i.e. the stellar-mass BHs), Agol \&
Kamionkowski (2002) used a best guess estimate of $\eta_{\rm X}\sim
10^{-5}$. } $\eta_{\rm X}=10^{-6}$.

For our galaxy model with mass $M=2\times 10^{12}M_\odot$, and accretion from the 
local ISM, we find (see Figure 9) that there are no sources with bolometric luminosity 
larger than $\sim 10^{39}$ erg/s, even for a large efficiency of $0.1$. 
Clearly, these sources cannot contribute to the ultra-luminous X-ray sources,
with $L_X\ga 10^{39}$ erg/s, observed in local galaxies. We find
the X-ray luminosity to be at most a few $\times 10^{38}$ erg/s (see Figure 10),
if accretion proceeds at high efficiency through an SS disc.     
We need to stress that, although we find some very massive slung black                     
holes at typical ISM densities (see Fig. 8), they have                    
rather large velocities, and this suppresses the accretion rate and prevents 
these sources from exhibiting very large X-ray luminosities
It should also be noted that, although the number of BHs is largest
in the halo and the smallest in the disc, the bright end of the
luminosity function is dominated by the BHs in the disc when accretion
is solely from the ISM. This is due to the typically larger densities
and lower temperatures of the disc.
If the accretion mode of these BHs were similar to the one in the 
Galactic Centre,
the bright end of X-ray luminosity would barely reach the $\sim 10^{34}$ erg/s level, 
several orders of magnitude below the Eddington limit for a solar mass BH. 

In our model in which local IMBHs are formed through mergers of smaller
BHs during the galaxy assembling process, bigger galaxies are more
likely to harbour more massive wandering BHs (compare Figures 3 and 4).
Given the scaling of the accretion rate with $M_{\rm BH}^2$, this
implies that, the larger the galaxy mass, the more luminous will be
the sources associated with accreting IMBHs. This is shown in Figure 11,
where the bolometric luminosity function is computed for a galaxy
of $M=10^{13}M_\odot$. Both a spiral and an elliptical galaxy models are considered.
If the accretion efficiency is high ($\eta\sim 0.1$), a spiral galaxy 
has a 10\% probability of hosting a source with $L_x\ga 10^{40}$ erg/s. 
This is however not the case for an elliptical galaxy, where the
lower ISM densities suppress the accretion emission. No ULXs due to IMBHs are  
expected even in a very massive elliptical galaxy and for a high accretion
efficiency.

\subsection{``Dressed'' wandering BHs}
All the previous results have been produced under the assumption that the BHs left over by minor 
mergers have been completely stripped of their envelope and accrete from their
surrounding ISM ({\em naked} BHs scenario). However, it is possible that, during the merger 
process, BHs are not completely stripped of their surrounding core of material.
Following tidal stripping, a bound core can remain, 
enveloping the central BH. Mayer \& Wadsley (2003) have performed
high-resolution N-Body/SPH simulations that follow the evolution of small disc satellites 
falling into the potential of a much larger halo. Tidal shocks remove most of the outer dark 
and baryonic mass. In addition, ram pressure affects the gaseous component, stripping it even more
severely. After a Hubble time, about $30\%$ of the stars have been stripped, 
and $80-90$\% of the gas disc has been removed, leaving a system resembling a gas-poor 
dwarf spheroidal. A reasonable value for the gas density in satellite remnants, is the 
typical gas density  $\rho\sim 10^{-3}{\rm cm}^{-3}$, 
found in dwarf galaxies of the 
local group (Grebel, Gallagher \& Harbeck 2003). The density scatter is large, varying by
about 3 orders of magnitude from one system to another. 
An optimistic upper limit for the gas densities was adopted by Islam et al. (2003b, c). 
Their estimate assumes that all the baryons in the satellite remnants are 
under the form of gas, and 
that the baryon content equals the cosmological baryon fraction. In this case, for a standard 
$\Lambda$CDM cosmology the gas density in the remnants is of order 
$\rho\sim 8\times 10^{-2}{\rm cm}^{-3}$.

Given the uncertainties involved in the estimates of the density of these
possible remnant cores, we prefer to just give some indicative results
within the context of this scenario. In particular, we find that about
10\% of the galaxies with mass $\sim 2\times 10^{12}M_\odot$ will have
a source with bolometric luminosity of $L_{\rm bol}\ga 1.5\times
10^{39}(\eta/0.1)(\rho/0.1\,{\rm cm}^{-3})$ erg/s for the 3$\sigma$
peaks, and $L_{\rm bol}\ga 2\times 10^{41}(\eta/0.1)(\rho/0.1\,{\rm cm}^{-3})$ 
for the 3.5$\sigma$ peaks.  It should also be noted that,
if the BHs do accrete from a remnant core, then the luminosity
function of the BHs will be independent of their location, and
therefore the most luminous ones are likely to be found in the halo,
where their number is the largest.

\section{Comparison with previous work}
Our results are much more pessimistic than previous investigations involving IMBHs remnants 
of the first stars (Islam, Taylor \& Silk 2003a,b,c). Although the hierarchical framework is 
similar this paper and the Islam et al. papers rely on different assumptions about the 
abundance of primordial seeds and about the dynamical and mass growth history of MBHs. 
First, their preferred model assumed seed holes, of $m_\bullet= 260\,\msun$, in 3-$\sigma$ 
peaks at $z=24.6$, leading to a number of seeds almost 3 orders of magnitude larger than that 
in our reference model (3.5-$\sigma$ peaks at $z=20$).  Second, their merger efficiency is 
much larger, since they assume that any MBH coming within $1\%$ of the virial radius of 
the host halo actually merges with the central BH. Finally, they ignore the mass growth 
of MBHs due to accretion of gas, in a luminous phase as quasars, thus requiring a large number 
of MBH mergers in order to reproduce the observed local $m_{\rm BH}-\sigma_*$ relation. 
In our model, instead, it is gas accretion, following galactic major mergers, which determines
the mass of SMBHs today, with mergers playing a secondary role, thus requiring a much smaller 
number of seeds to reproduce the observed SMBHs masses. 

As a result, our predictions for the number of X-ray sources in a galaxy
of given mass generally fall below the predictions of Islam et al. (2003c),
when similar estimates for the density and temperature of the core remnant
are adopted\footnote{Note that we consider the possibility of BHs 
accreting from a baryonic
remnant only for the satellites, whose mass distribution
has an upper end cutoff which is smaller than that of the sling BHs (see Fig. 3 and 4).}.  

\section{Summary  and discussion}
In this paper
we have studied the dynamical evolution with cosmic time of a population 
of IMBHs wandering in galactic halos, in the context of popular hierarchical structure 
formation theories. We follow the merger history of dark matter 
halos and associated SMBHs via cosmological Monte Carlo realizations of the 
merger hierarchy from the end of the dark ages to the present in a $\Lambda$CDM cosmology.
Massive black holes form as the end-product of the first generation of stars and then evolve 
in a hierarchical fashion following the fate of their host halos. Along the merger history
of dark matter halos, the BHs embedded in their inner regions sink to the centre of coalescing
larger and larger galactic structures owing to dynamical friction, 
accrete a fraction of the gas in the merger remnant to become supermassive,
and form a binary system. Stellar dynamical processes drive the binary to 
harden and eventually coalesce. The dynamical evolution of SMBH binaries may be disturbed 
by a third  incoming BH, if another major merger takes place before the pre-existing binary has 
had time to coalesce. The three BHs are likely to undergo a 
complicated resonance scattering interaction, leading to the recoil of the binary and
of the single BH, thus creating a population of wandering BHs. Other coalescing binary BHs 
become wandering due to  the non-zero net linear momentum carried away by gravitational waves 
if a binary is asymmetric. BHs retained within the potential well of the host halo are then
bound to slowly travel back towards the galactic centre under the effect of dynamical friction.
The population of wandering BHs comprises also another class of holes, those 
originally hosted by halos which are
small compared to the primary halo. In this case the
wandering holes can, or not,  be still surrounded by the remnants of their old host.

Our simulations have allowed us to estimate the number of black holes,
their mass distribution and position within the galaxy.  The total
number of wandering IMBHs correlates with the mass of the galaxy at
$z=0$, and the BHs have typically larger masses.  As a result, the
number of point sources due to IMBHs is also expected to correlate
with the mass of the galaxy, and so the luminosity of the brightest
sources.  To zeroth order, no correlation is expected in our scenario
of IMBHs formation and the star formation rate (SFR) in the galaxy at
$z=0$, as the mass function and positions of the BHs within the galaxy
are essentially determined by its past merger history.  However, if a
galaxy has a high SFR, then it is likely to also have a larger filling
fraction of dense molecular clouds, which would significantly enhance
the accretion rate if a BH happened to wander inside them.   In the
scenario recently proposed by Krolik (2004), BHs can be formed from
primordial abundance gas or from stellar mergers \footnote{For studies of stellar merging 
see also Portegies Zwart \& McMillan (2002) and Gurkan, Rasio \& Freitag (2004);            
note that the formation of IMBHs by runaway                                              
stellar merging is possible only in young, compact clusters.}, and these mechanisms
would make them to preferentially populate the disk of galaxies,
therefore making them brighter, especially if they move for most of
the time inside dense regions. 

In our scenario of BH formation in mini-halos and growth through
subsequent galaxy merging, we find that, if the wandering IMBHs do not
carry any substantial baryonic core with them, then a Milky Way sized
galaxy is not likely  to harbour sources with $L\ga 10^{39}$
erg/s. On the other hand, about 10\% of spiral galaxies with mass
$\sim 10^{13}M_\odot$ would likely have a source with luminosity $\ga
10^{40}$ erg/s.  The brightest sources are expected to be found in the
disk, due to its typical lower temperature and larger density than the
bulge and the halo. No ULXs are expected in ellipticals, even of large
mass ($\sim 10^{13}M_\odot$), due to their average lower density and the lack of a cooler,
dense disk which boosts the emission.
At sub-Eddington luminosities, the luminosity function of the
naked IMBH population in a typical, MW-size galaxy is well 
below the observed point source population in {\em Chandra} surveys
(Colbert et al. 2004). This is not surprising, as this point source
population in the sub-Eddington range is dominated by high mass
X-ray binaries (Grimm et al. 2004). 

Our results imply that, in order to produce ULXs from IMBHs in MW-type
galaxies (and ellipticals for that matter), it is necessary that the
wandering IMBHs are not completely stripped of their baryonic core,
and this remnant material feeds the accretion at a rate higher than
what would be allowed from the ISM alone. Even allowing an high
efficiency of $\sim 0.1$, a density $\ga 0.01 {\rm cm}^{-3}$ is required in order for
a few percent of MW-type galaxies to host a source with luminosity
$\ga 10^{40}$ erg/s.  However, the density cannot be significantly higher than
that. In our models, the high tail of the mass distribution of IMBHs
extends to $\sim 10^6-10^8 M_\odot$, and the observational lack of
sources with luminosities $\ga 10^{44}$ erg/s implies that these IMBHs
must accrete at sub-Eddington rates.

In this paper, however, we have investigated the luminosity that IMBHs can attain
in normal, quiescent galaxies. Starbursting galaxies, such as the Antennae 
(Zezas \& Fabbiano 2002) or the 
Cartwheel galaxy (Gao et al 2003) show much larger population of ULXs compared to quiescent 
galaxies of the same size. In particular, in the Cartwheel galaxy these sources are found in the 
outer star-forming ring, believed to be created by radially expanding density waves caused by
a plunging merger. Note that all these sources are at a distance larger than $\sim 10$ kpc 
from the centre of the galaxy, so in our scheme for a quiescent spiral, they would be found
in the low-density, gas-poor galactic halo. The density waves expanding outward, though, 
provide a temporary high density environment that may supply fuel for an IMBH to accrete and 
shine at high luminosities, higher also than those predicted for IMBHs travelling in galactic discs.

\section*{Acknowledgments}
We have benefited from discussions with M. Colpi, T. J. Cox, P. Madau and C. Nipoti.

\label{lastpage}


\begin{thebibliography}{99}
\bibitem{ak} Agol, E. \& Kamionkowski, M. 2002, MNRAS, 334, 553
\bibitem{b0} Armitage, P. J., \& Natarajan, P. 2002, ApJ, 567, L9
\bibitem{b1}  Baganoff F. K. et al., 2001, Nat., 413, 45
\bibitem{b100} Barnes J. E. 1988, ApJ, 331, 699
\bibitem{b2}  Begelman M. C., 2002, ApJ, 568, L97
\bibitem{b200} Begelman M. C., Blandford R. D., Rees M. J. 1980, Nature, 287, 307
\bibitem{b3}  Begelman, M.~C., Celotti, A., MNRAS in press, astro-ph/0407063
\bibitem{bell} Bell E.~F. \&  de Jong R.~S. 2000, MNRAS, 312, 497
\bibitem{bt} Binney, J., \& Tremaine, S. 1987, Galactic Dynamics (Princeton: Princeton Univ. Press)
\bibitem{blaes} Blaes, O., Warren, O. \& Madau, P. 1995, apJ, 454, 370.
\bibitem{b300} Blandford R.~D. \& Rees M.~J. 1974, MNRAS, 169, 395
\bibitem{b4}  Blandford R.~D., Begelman M.~C., 1999, MNRAS, 303, L1
\bibitem{b5}  Bland-Hawthorn J., Reynolds R., 2000, Encyclopedia of Astronomy \& Astrophysics, MacMillan and Institute of Physics Publishing
\bibitem{b6}  Bondi H., Hoyle F., 1944, MNRAS, 104, 273
\bibitem{b7}  Colbert E.~J.~M., Mushotzky R. F., 1999, ApJ, 519, 89
\bibitem{b8}  Colbert E.~J.~M., Heckman T.~M., Ptak A. F., Strickland D.~K.,Weaver K.~A. 2004, ApJ, 602, 231
\bibitem{b9}  Cropper M. et al., 2004, MNRAS, 349, 39
\bibitem{b99} Damour T. 2001, PhRvD, 64l4013D
\bibitem{b90} Escala, A., Larson, R. B., Coppi, P. S., \& Mardones, D. 2004, ApJ, 607, 765
\bibitem{b10} Fabbiano G., 1989, AR\&A, 27, 87
\bibitem{bb10} Favata, M., Hughes, S. A., \& Holz, D. E. 2004, ApJ, 607, L5  
\bibitem{b101} Ferrarese L., \& Merritt, D. 2000, ApJ, 539, L9
\bibitem{b102} Ferrarese L. 2002, ApJ, in press (astro-ph/0203469)
\bibitem{b501} Fitchett, M. 1983, MNRAS, 203, 1049
\bibitem{b503} {Fukugita} M., {Hogan} C.~J. \& {Peebles} P.~J.~E. 1998, ApJ, 503, 518
\bibitem{b803} Gao Y., Wang Q., Daniel A. P. N. \& Lucas R. A. 2003, ApJL, 596, 171
\bibitem{b103} {Gebhardt} K., \etal 2000, ApJ, 543, L5
\bibitem{grebel} Grebel E. K.; Gallagher J. S. III \& Harbeck, D. 2003, AJ, 125, 1926
\bibitem{b11} Grimm H.-J., Gilfanov M., Sunyaev R., 2003, MNRAS, 339, 793
\bibitem{ato} {G{\" u}rkan}, M.~A., {Freitag}, M. \& {Rasio}, F.~A. 2004, ApJ, 604, 632
\bibitem{b662} Hernquist L. 1990, ApJ, 356, 359
\bibitem{b110} {{Hills}, J.~G.}, \& Fullerton, L.~W. 1980, AJ, 85, 1281
\bibitem{b111} Hut, P., \& Rees, M.~J. 1992, {MNRAS}, 259, 27
\bibitem{i1} {Islam} R.~R., {Taylor} J.~E. \& {Silk}, J. 2003a, MNRAS, 340, 647
\bibitem{i2} {Islam} R.~R., {Taylor} J.~E. \& {Silk}, 2003b, astro-ph/0307171
\bibitem{i3} {Islam} R.~R., {Taylor} J.~E. \& {Silk}, 2003c, astro-ph/0309558
\bibitem{b12} King A.~R., Davies M.~B., Ward M.~J., Fabbiano G.,Elvis M., 2001, ApJ, 552, L109
\bibitem{b13} Krolik J., 2004, ApJ in press, astro-ph/0407285
\bibitem{b133} Madau, P., \& Rees, M. J. 2001, ApJ, 551, L27 
\bibitem{b1333}Madau P., Quataert E. 2004, ApJL, 606, 17
\bibitem{b131} {Magorrian} J., \etal 1998, AJ, 115, 2285
\bibitem{b728}{Mathews} W.~G. \& {Brighenti}  F. 2003, ARA\&A, 41, 191
\bibitem{mw} Mayer L. \& Wadsley J. 2003, astro-ph/0309073 
\bibitem{merr} {Merritt}, D. 2000, in ASP Conf. 197, Dynamics of Galaxies: from the 
Early Universe to the Present, ed. F. Combes, G. A. Mamon, \& V. 
Charmandaris (San Francisco:ASP), 221
\bibitem{b1383} Merritt, D., Milosavljevic, M., Favata, M., Hughes, S. A., \& Holz, D. E. 
2004, ApJ, 607, L9
\bibitem{c132} Mikkola, S., \& Valtonen, M. J. 1990, ApJ, 348, 412
\bibitem{b14} Miller M.~C. , Hamilton D. P., 2002, ApJL, 570, 69
\bibitem{b140} Milosavljevic, M., \& Merritt, D. 2001, ApJ, 563, 34
\bibitem{b15} Narayan R., Yi I., 1995a, ApJ, 452, 710
\bibitem{b16} Narayan R., Yi I., 1995b, ApJ, 444, 231
\bibitem{b160} Navarro, J. F., Frenk, C. S., \& White, S. D. M. 1997, ApJ, 490, 493
\bibitem{b166} Nipoti C., Stiavelli M., Ciotti, L., Treu, T., Rosati, P. 2003, MNRAS, 344, 748
\bibitem{b17} Perna R., Narayan R., Rybicki G., Stella L., Treves, A. 2003, ApJ, 594, 936
\bibitem{b18} Perna R., Stella, L. 2004, ApJ, 615, 222
\bibitem{pz} Portegies Zwart, S. F. \& McMillan, S. L. W. 2002, ApJ, 576, 899
\bibitem{b181} {{Quinlan}, G.~D.} 1996, {NewA}, 1, 35
\bibitem{b180} {Quinlan} G.~D., \& {Hernquist} L. 1997, {NewA}, 2, 533
\bibitem{b19} Rees M.~J., Phinney E.~S., Begelman M.~C., Blandford R.~D., 1982, Nat,  295, 17
\bibitem{b190} Saslaw W. C., Valtonen M. J., \& Aarseth, S. J. 1974, ApJ, 190, 253
\bibitem{b20} Shakura N.~I., Sunyaev R.A., 1973, A\&A, 24, 337
\bibitem{b503} Shen S. et al. 2003, MNRAS, 343, 978
\bibitem{b205} Springel V. \& White S.~D.~M. 1999, MNRAS, 307, 162
\bibitem{b2001}Taffoni G., Mayer L., Colpi M. \& Governato F. 2003, MNRAS, 341, 434
\bibitem{b2000} Tormen, G. 1997, {MNRAS}, 290, 411
\bibitem{b220} {Valtonen}, M.~J., \& {Hein{\" a}m{\" a}ki}, P. 2000, ApJ, 530, 107
\bibitem{b222} van den Bosch, F. C., Lewis, G. F., Lake, G., \& Stadel, J. 1999, ApJ, 515, 50 
\bibitem{b21} Volonteri M., Haardt F., \& Madau P. 2003, ApJ, 582, 559 (VHM03)
\bibitem{b22} Volonteri M., Madau, P. \& Haardt, F. 2003, ApJ, 593, 661
\bibitem{b225} White, S. D. M.  1980, MNRAS, 191, 1
\bibitem{xu} Xu, G., \& Ostriker, J. P. 1994, ApJ, 437, 184
\bibitem{b23} Yu, Q. 2002, MNRAS, 331, 935
\bibitem{b24} Zezas A.\& Fabbiano G.  2002, ApJ, 577, 726
\end{thebibliography}
\end{document}